\documentclass[conference]{IEEEtran}
\IEEEoverridecommandlockouts

\usepackage{graphicx}
\usepackage{amsmath}
\usepackage{amssymb}
\usepackage{enumerate}
\usepackage{url}
\usepackage{bm}
\usepackage{framed}
\usepackage{mathptmx}
\usepackage{lettrine}
\usepackage{eucal}
\usepackage{nicefrac}
\usepackage{listings}
\usepackage{adjustbox}
\usepackage{multirow}
\usepackage{multicol}
\usepackage{tikz}

\newtheorem{prop}{Proposition}

\newtheorem{lem}{Lemma}

\newtheorem{definition}{Definition}

\usetikzlibrary{shapes.geometric, arrows}
\tikzstyle{startstop} = [rectangle,  minimum width=3cm, minimum height=1cm,text centered,text width=10cm, draw=black ,fill=gray!20]
\tikzstyle{process} = [rectangle, minimum width=3cm, minimum height=1cm, text centered,text width=10cm, draw=black,fill=orange!20]
\tikzstyle{arrow} = [thick,->,>=stealth]
\tikzstyle{state}=[shape=circle,draw=blue!50,fill=blue!20]
\tikzstyle{observation}=[shape=rectangle,draw=orange!50,fill=orange!20]
\tikzstyle{lightedge}=[<-,dotted]
\tikzstyle{mainstate}=[state,thick]
\tikzstyle{mainedge}=[<-,thick]
\title{Optimizing Age-of-Information in Adversarial Environments with Channel State Information}
\author{%
\IEEEauthorblockN{Avijit Mandal\IEEEauthorrefmark{1}, Rajarshi Bhattacharjee \IEEEauthorrefmark{2}, Abhishek Sinha\IEEEauthorrefmark{3}}
  \IEEEauthorblockA{\IEEEauthorrefmark{1}\IEEEauthorrefmark{3}Dept. of Electrical Engineering, Indian Institute of Technology Madras,
                    Chennai 600036, India\\ 
  \IEEEauthorrefmark{2} College of Information and Computer Sciences, University of Massachusetts at Amherst,             \\  }
  Email:  
   \IEEEauthorrefmark{1} avijitbesu1995@gmail.com, 
   \IEEEauthorrefmark{2}rbhattacharj@umass.edu, 
  \IEEEauthorrefmark{3} abhishek.sinha@ee.iitm.ac.in\\
\thanks{\IEEEauthorrefmark{1} \IEEEauthorrefmark{2} Work done while the first and second authors were serving as project associates at the Indian Institute of Technology Madras.}
                   
}
\begin{document}
\maketitle
\begin{abstract}
This paper considers a multi-user downlink scheduling problem with access to the channel state information at the transmitter (CSIT) to minimize the Age-of-Information (AoI) in a non-stationary environment. The non-stationary environment is modelled using a novel adversarial framework.  In this setting, we propose a greedy scheduling policy, called \texttt{MA-CSIT}, that takes into account the current channel state information. We establish a finite upper bound on the competitive ratio achieved by the \texttt{MA-CSIT} policy for a small number of users and show that the proposed policy has a better performance guarantee than a recently proposed greedy scheduler that operates without CSIT. In particular, we show that access to the additional channel state information improves the competitive ratio from $8$ to $2$ in the two-user case and from $18$ to $\nicefrac{8}{3}$ in the three-user case. 
 Finally, we carry out extensive numerical simulations to quantify the advantage of knowing CSIT in order to minimize the Age-of-Information for an arbitrary number of users. 
\end{abstract}

\section{Introduction}

\lettrine{I}{n} addition to throughput, delay, and spectral efficiency, the \emph{Age-of-Information} (AoI) metric has recently emerged as one of the key determinants of the  Quality of Service (QoS) offered by the next-generation wireless networks. The AoI metric, first introduced in \cite{kaul2012real}, measures the \emph{freshness} of information available to the users in real-time. Ever since the pioneering work by Kaul et al., there has been an extensive body of work on optimizing and understanding the design implications of AoI in communication systems. See \cite{sun2019age} for a comprehensive introduction to the recent advances in this area. In order to keep the analysis tractable, most of the existing papers on AoI assume stationary stochastic system models \cite{AoI_infocom, AoI_ToN}. Furthermore, the usual performance guarantees given in the literature in connection with AoI are almost always asymptotic in nature. On the contrary, applications where the AoI metric is critical to the system performance, such as the ultra-reliable low latency communication (URLLC) and mission-critical communication in cyber-physical systems, typically operate far from the stationary regime \cite{park2020extreme}. For acceptable performance, these applications also require stringent non-asymptotic upper limits on the age-of-information. To address this issue, in this paper, we focus on designing robust scheduling algorithms that ensure the maximum information freshness for the end-users, \emph{irrespective} of the possibly time-varying statistics of the underlying wireless channel. In our recent papers \cite{ISIT2020Sinha,SinhaMAMA2020,sinha2020optimizing}, we introduced an adversarial version of Binary Erasure Channel (BEC) model, and showed that a greedy scheduling policy is approximately  competitively optimal. These papers assume that the channel states are adversarially chosen and the scheduler \emph{does not} have access to the current channel state information (CSIT). In the present paper, we extend our previous results to the setting where the channel state information of the current slot is available to the scheduler. The main objective of this paper is to quantify the provable improvement in performance due to the availability of CSIT compared to the setting when the transmitter is oblivious to the current channel state. Due to the complexity of the analysis, we only have been able to theoretically analyze the setting when the number of users ($N)$ is either two or three. Our numerical experiments suggest the AoI advantage continues in the presence of CSIT even when the number of users is large. We anticipate that the tools and techniques developed in this paper will be useful to tackle the general problem with an arbitrary number of users.  
In this paper, we claim the following two main contributions:
\begin{enumerate}
\item 	For the adversarial channel model, we establish an improved upper bound on the competitive ratio for a greedy online scheduling policy that has access to the current CSIT. We show that the proposed online policy is 2-competitive when $N=2$ and  $\nicefrac{8}{3}\sim 2.67$-competitive when $N=3$. This improves the previously known tight upper-bounds on the competitive ratios (without CSIT), which are known to be $8$ (for $N=2$) and $18$ (for $N=3$) respectively (see Theorem 3 of \cite{banerjee2020fundamental} and Theorem 1 of \cite{sinha2020optimizing}, where the competitive ratio is bounded by $2N^2$ for any $N \geq 1$).   
\item We numerically compare the performance of the online scheduling policy which knows channel states at the current slot with a greedy  online scheduling policy which does not have the current channel state information. Our results show that the AoI is substantially reduced with CSIT. 

\end{enumerate}

The rest of the paper is organized as follows. In Section \ref{model}, we describe our adversarial system model and formulate the problem. In Section \ref{analysis}, we derive the competitive ratio of a greedy scheduling policy for the case $N=2$ and $N=3$. In Section \ref{sims}, we present our simulation results, and finally, in Section \ref{conclusion}, we conclude the paper with a brief discussion on possible future research directions.

\section{System Model and Problem Formulation} \label{model}
We consider an online scheduling problem with $N$ users located in the coverage area of a single Base Station (henceforth referred to as BS). Time is slotted, and at the beginning of every slot, a fresh update packet arrives at the BS from some external source. Such traffic models are known as the saturated traffic models in the literature \cite{borst2001dynamic, holtzman2001asymptotic, kushner2004convergence}. Each of the $N$ users are interested in receiving the fresh packet at each slot to keep up-to-date with the external source. Once a fresh packet arrives, the BS beamforms and schedules a packet transmission to one of the $N$ users according to a scheduling policy $\pi$. The downlink channels from the BS to the users are assumed to be non-stationary, modeled as an adversarial binary erasure channel, whose states are dictated by an adversary. In particular, the downlink channel state for any user could be either \textsf{Good}
or \textsf{Bad} as determined by the adversary. The online scheduling policy $\pi$, equipped with the channel state information 
(CSIT), knows the current channel states of all users before the scheduling decision for a slot is made.
Making use of the current channel state information, the policy selects a user having a \textsf{Good} channel (if any) and then transmits
the latest packet from the BS to the user. The adversary controlling the channel states may know the scheduling policy as well. This adversarial framework was first introduced in our recent papers \cite{ISIT2020Sinha,SinhaMAMA2020,sinha2020optimizing}.

\begin{figure}[ht]
\centering

\includegraphics[scale=0.4]{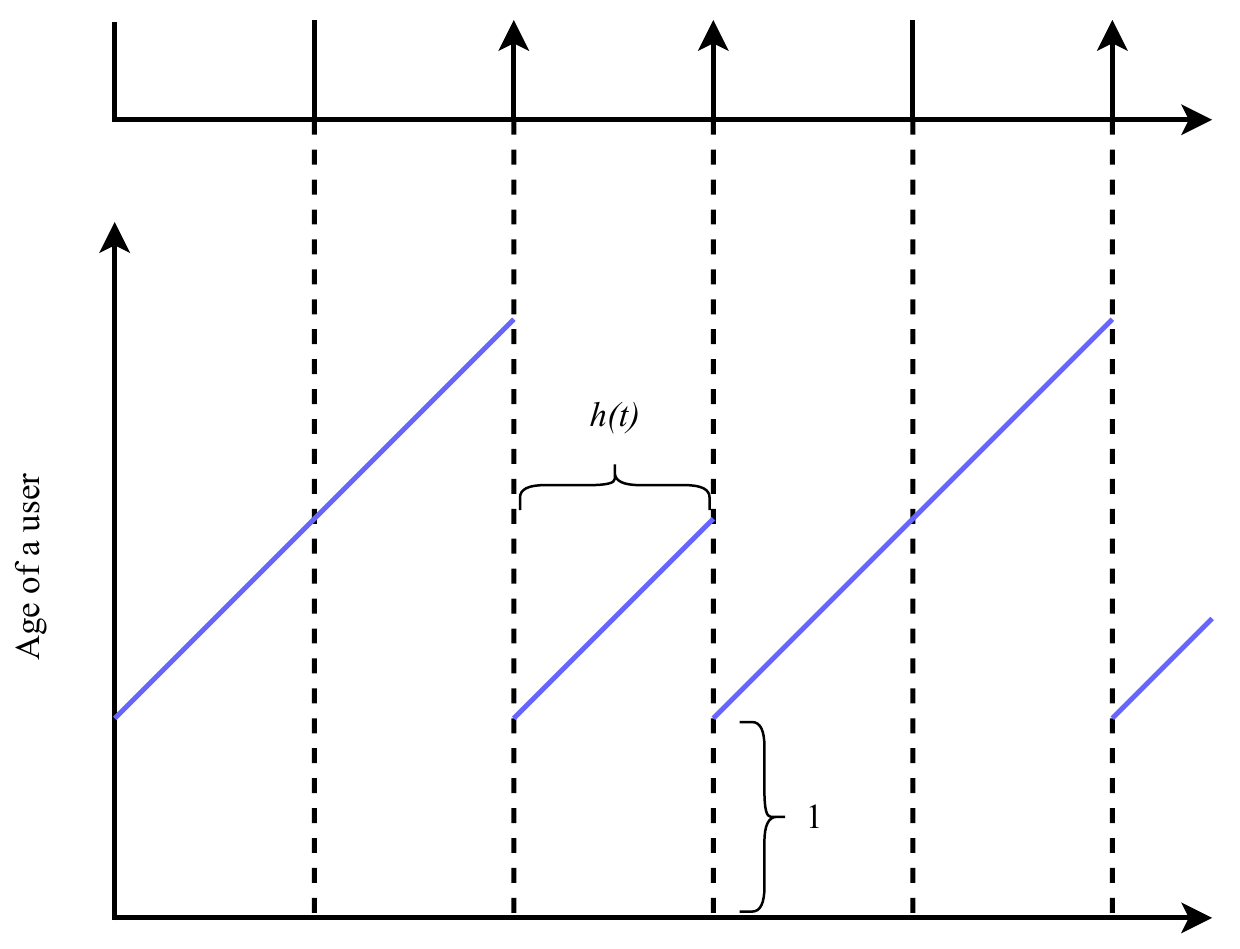}  
\caption{Time evolution of $h_{i}(t)$, an arrow indicates successful transmission of fresh packet to the $i^{th}$ user.}\label{timeevol}
\end{figure}

Our objective is to design a scheduling policy that competitively optimizes the average freshness of information for all the users. For any time slot $t\geq 1$, let $t_{i}(t)$ denotes the last time slot when the $i^{th}$ user successfully received a packet from the BS. The \emph{Age of Information} (AoI) for the $i^{th}$ user at slot $t$ is defined as:
\begin{equation*}
    h_{i}(t)= t-t_{i}(t).
\end{equation*}
    In other words, the quantity $h_{i}(t)$ measures the number of time slots before which the $i^{th}$ user received the last packet prior to time $t$. The $N$ dimensional vector $\bm{h}(t)$ represents the collection of  AoI for $N$ users at time $t$ where $i^{th}$ element of the vector refers to the AoI of the $i^{th}$ user \emph{i.e.} $h_{i}(t)$. The age $h_{i}(t)$ increases linearly with time until the $i^{th}$ user receives a fresh packet. Once a user receives a fresh packet, its AoI instantaneously drops to unity. See Fig.~\ref{timeevol} for an illustration of the evolution of AoI.
    \paragraph*{Objective function} Throughout this paper, we consider optimizing the  total AoI metric, which is defined as the sum of  AoI cost incurred for all users over the entire time horizon under consideration.
    Hence, the objective function for the AoI minimization can be expressed as:
 \begin{equation}
     \textrm{AoI}(T)=\sum_{t=1}^{T}(\sum_{i=1}^{T}h_{i}(t))
 \end{equation}

\subsection{Performance metric}
To quantify the performance of any scheduling policy, we use the standard notion of competitive ratio from the online algorithms literature \cite{fiat1998,albers1996competitive}. To be specific, the competitive ratio of any online policy $\pi$ is defined as the ratio between worst-case cost incurred by the policy $\pi$ and the cost incurred by the offline clairvoyant optimal policy \textsf{OPT}, which knows the entire sequence of channel states in advance. Let $\sigma$ denote any sequence of channel states. The competitive ratio of an online policy $\pi$ can be expressed as:
\begin{equation}
    \eta^{\pi}=\sup_{\sigma}\Bigg(\frac{\text{Cost of the policy $\pi$ on $\sigma$}}{\text{Cost of \textsf{OPT} on $\sigma$}} \Bigg),
\end{equation}
where, in the above, the supremum is taken over all possible finite length channel state sequences $\sigma$. 

\subsection{Scheduling Policies}
In this paper we analyze the performance of the following online policy:
\subsubsection*{Max-Age with CSIT policy (\textsf{MA-CSIT})}
At each time slot, the scheduler determines the current channel states of all users using the CSIT. The BS then schedules a fresh packet transmission to the user having the \emph{highest age} among all users currently having a \textsf{Good} channel.
If at any time slot, no channel is in \textsf{Good} state, the \textsf{MA-CSIT} policy does not schedule a packet transmission to any user.

\begin{figure}[ht]
\centering
\includegraphics[scale=0.4]{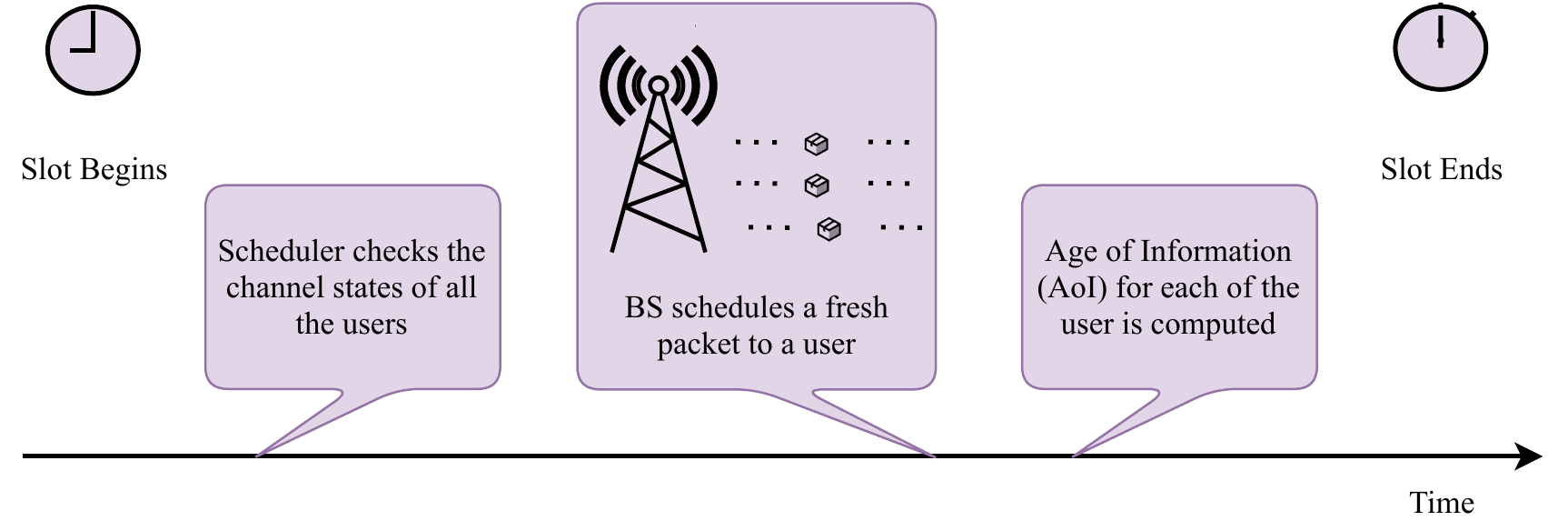}  
\caption{Timeline of \textsf{MA-CSIT} policy at a particular time slot}\label{timeline}
\end{figure}

For bounding the competitive ratio of the \textsf{MA-CSIT} policy, we need to characterize the \emph{Offline} optimal policy (\textsf{OPT}). The \textsf{OPT} policy is assumed to know the channel states of all the users for the entire time duration \emph{a priori}. Hence, the performance of any other scheduling policy is dominated by that of \textsf{OPT}. However, the \textsf{OPT} policy can not be implemented in an online fashion as it assumes the knowledge of the future channel states.  

\paragraph*{Baseline} In order to determine the benefit of having CSIT, we compare the performance of the \textsf{MA-CSIT} policy with the Max-Age policy that does not consider the channel state information \cite{ISIT2020Sinha,sinha2020optimizing}.  Under the Max-Age policy, at each time slot $t$, BS schedules a fresh packet to the user which has the highest age among all the users, irrespective of the current channel states. Hence, if the channel state at the scheduled user-end turns out to be \textsf{Bad}, the packet is lost.  
\section{Performance Analysis} \label{analysis}
In this section, we bound the competitive ratio of the \textsf{MA-CSIT} policy from the above. 
\paragraph*{A note on determining the Max-age users}
To begin with, at any time slot, we first sort the users according to descending order of their ages under the \textsf{MA-CSIT} policy. The user, who has the highest age among all the users (under the \textsf{MA-CSIT} policy) is called the \emph{Max-age} user (ties are broken arbitrarily).  Similarly, the $m^{th}$ user in the sorted list is called the $m^{th}$ \emph{Max-age} user. Thus the \emph{Max-age} user corresponds to $m=1$ in the sorted list at that time slot. Naturally, under a different policy (\emph{e.g.,} \textsf{OPT}) the \emph{Max-Age} user may not have the highest age among all users. 

Next, we recall the concept of a time \emph{interval}, first introduced in \cite{banerjee2020fundamental},  
\begin{framed}
\begin{definition}(Interval)
A new interval is said to begin when the \emph{Max-age} user  transmits a packet successfully under the \textsf{MA-CSIT} policy.
\end{definition}

\end{framed}
Hence, an interval continues until the channel corresponding to the \emph{Max-age} user becomes $Good$.
Let  the quantity $h_{k}^{i,t}$ denote the age of the $k^{th}$ user\footnote{here $k$ refers to the index of the user which is same for both \textsf{MA-CSIT} and \textsf{OPT} policy for the entire time duration $T$. Please note that, this $k$ does not refer to the index of the user in the sorted list which is prepared at every time-slot to determine the ordering of the Max-age users on the basis of the ages of the users under the \textsf{MA-CSIT} policy. For example the user 1 at a certain time-slot may become the Max-age user and at another time-slot may become $2$\textsuperscript{nd} Max-age user and so on, but its index remains same for the time duration $T$ for both the policies.} at the $t$\textsuperscript{th} time slot of the $i$\textsuperscript{th} interval under the \textsf{MA-CSIT} policy. Also, let $o_{k}^{i,t}$ denote the age of the $k^{th}$ user at the $t^{th}$ time slot of $i^{th}$ interval under \textsf{OPT} policy.  So $h_{k}^{i,1}$ denotes the age of the $k^{th}$ user at the first time slot of the $i^{th}$ interval under the \textsf{MA-CSIT} policy. The length of the $i^{th}$ interval is denoted as $I_{i}$ and the total AoI cost incurred by the \textsf{MA-CSIT} and the \textsf{OPT} policies on the $i$\textsuperscript{th} interval are denoted by $C_{\textsf{MA-CSIT}}(I_{i})$ and $C_{\textsf{OPT}}(I_{i})$ respectively.  With the above definitions in place, we now proceed to bound the competitive ratio of the \textsf{MA-CSIT} policy. 
\subsection{Competitive ratio of the \textsf{MA-CSIT} policy for $N=2$ users}\label{2user-upper bound}
\begin{framed}
\begin{prop}
The competitive ratio of  the \textsf{MA-CSIT} policy for $N=2$ users is upper bounded as $\eta^{\textsf{MA-CSIT}}\leq 2$.
\end{prop}
    
\end{framed}

\begin{figure}
\centering

\includegraphics[scale=0.4]{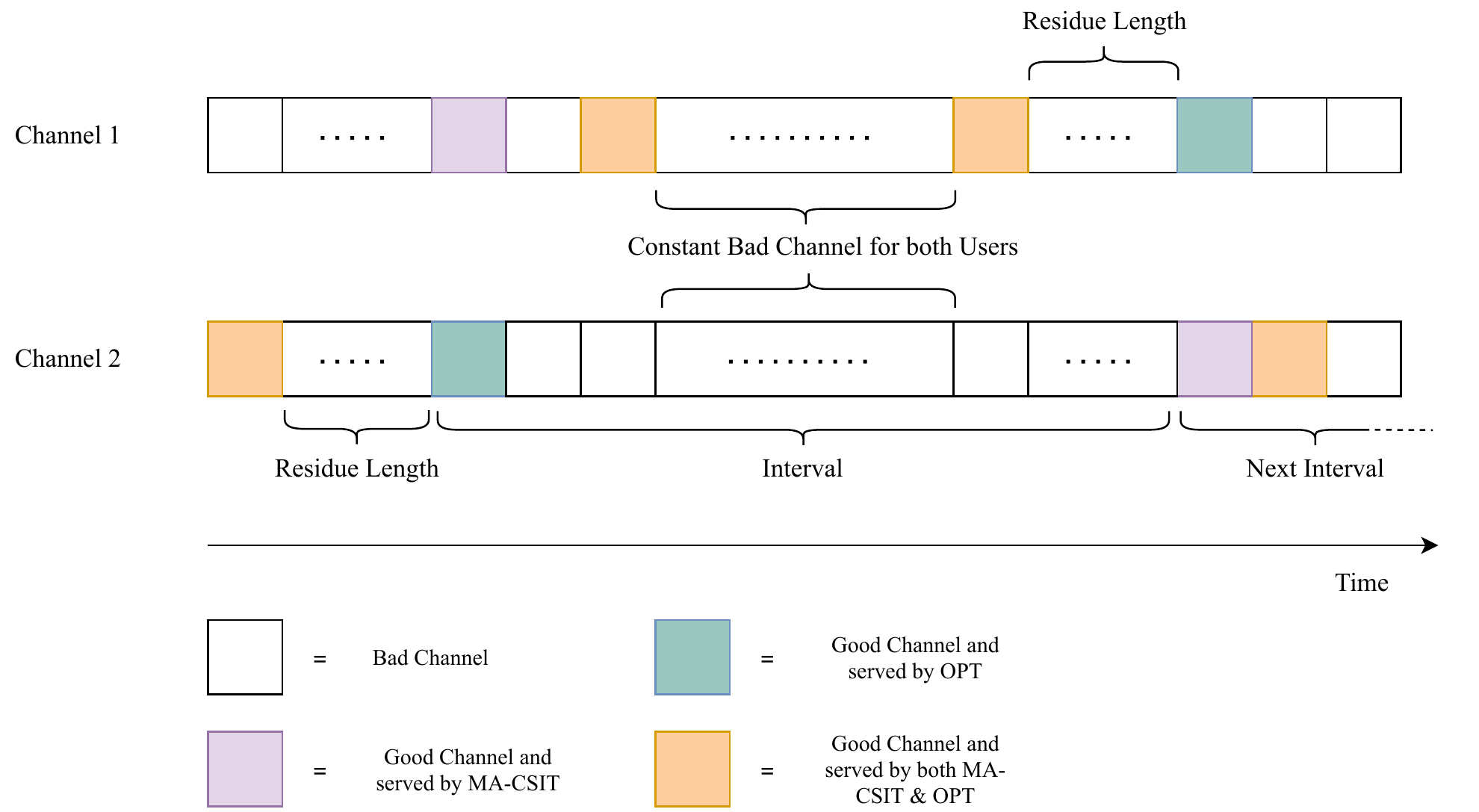}  
\caption{Illustration of residue length and interval construction for 2 users. Channel $i$ refers to the state of the channel corresponding to user $i$.}\label{up2}
\end{figure}

\begin{IEEEproof}
For two users, we can express the difference between the costs incurred by the \textsf{MA-CSIT} policy and \textsf{OPT} as:
\begin{equation}\label{n2}
    C_{\textsf{MA-CSIT}}(I_{i})-C_{\textsf{OPT}}(I_{i})=\sum_{t}(h_{1}^{i,t}-o_{1}^{i,t})+ \sum_{t}(h_{2}^{i,t}-o_{2}^{i,t}),
\end{equation}
where the index in the summation ranges over all slots in the $i$\textsuperscript{th} interval.
We now establish the following Lemma.
 \begin{framed} 
 \begin{lem}
  For the \emph{Max-age} user,  the age difference between the \textsf{MA-CSIT} policy and the \textsf{OPT} policy for every time slot $t$
 of $i^{th}$ interval remains constant. For example, if at the $i^{th}$ interval the user 1 remains the Max-age user then the age difference $(h_{1}^{i,t}-o_{1}^{i,t})$  remains constant throughout the interval $i$.
 \end{lem}
 \end{framed}
\begin{IEEEproof}
 Without any loss of generality, let us assume that the \textsf{MA-CSIT} policy serves the user 2 at the beginning of the $i^{th}$ interval, and at the $i^{th}$ interval, the user 1 becomes the \emph{Max-age} user.\\
We establish Lemma 1 on the basis of the following observation.
  Both \textsf{MA-CSIT} and \textsf{OPT} policies can not serve the \emph{Max-age} user until the channel corresponding to that user becomes \textsf{Good}. Furthermore, whenever the channel becomes $Good$, the \textsf{MA-CSIT} policy will serve the \emph{Max-age} user immediately and a new interval begins. Thus, within any interval, both the quantities $h_1^{i,t}$ and $o_1^{i,t}$ increase linearly. Hence,
 \begin{equation}
     h_{1}^{i,t}-o_{1}^{i,t}=h_{1}^{i,1}-o_{1}^{i,1}\quad \forall t. 
 \end{equation}
 \end{IEEEproof}
  Since we assume that user 1 is the Max-age user at the $i^{th}$ interval, we have $h_{1}^{i,1}>h_{2}^{i,1}$. The next interval, \emph{i.e.,} the $(i+1)^{th}$ interval begins when the \textsf{MA-CSIT} policy serves user 1. Thus we have,
  \begin{equation}
      C_{\textsf{MA-CSIT}}(I_{i})=C_{\textsf{OPT}}(I_{i})+(h_{1}^{i,1}-o_{1}^{i,1})I_{i}+\sum_{t}(h_{2}^{i,t}-o_{2}^{i,t})
  \end{equation}
   We now establish the following useful result. 
 \begin{framed}
  \begin{lem}
  For the user other than the \emph{Max-age} user, the age difference between the \textsf{MA-CSIT} and the \textsf{OPT} policy is always non-positive (\emph{i.e.,} $h_{2}^{i,t}-o_{2}^{i,t}\leq 0, \forall t$  for this case). 
 \end{lem}
 
 \end{framed}
 \begin{IEEEproof}
 To prove $h_{2}^{i,t}-o_{2}^{i,t}\leq 0$ $\forall t$ we use the following facts.  At the next time slots of the $i^{th}$ interval  whenever the channel corresponding to user 2 becomes $Good$, both the \textsf{MA-CSIT} and the \textsf{OPT} policies serve the user 2. The only scenario when the age of user 2 under the \textsf{MA-CSIT} policy becomes greater than age of user 2 under \textsf{OPT} \emph{i.e.} $h_{2}^{i,t}>o_{2}^{i,t}$ is when the channel corresponding to user 2 becomes $Good$ and \textsf{OPT} serves the user 2 but \textsf{MA-CSIT} does not. In other words the \textsf{OPT} policy serves the user 2 while the \textsf{MA-CSIT} policy serves the  user 1. Since we considered user 1 as the \emph{Max-age} user and if the \textsf{MA-CSIT} policy serves the user 1, from the definition of interval  the next interval \emph{i.e.} $(i+1)^{th}$ interval starts. 
 This implies at the $i^{th}$ interval the age of the user 2 under \textsf{MA-CSIT} will never become  more than the age under \textsf{OPT}. Hence,
   \begin{equation}\label{nrem2}
       h_{2}^{i,t}-o_{2}^{i,t}\leq 0 \quad \forall t
   \end{equation}
 \end{IEEEproof} 
 Combining the above two Lemmas, equation \eqref{n2} may be simplified as
 \begin{equation} \label{int_ineq}
     C_{\textsf{MA-CSIT}}(I_{i})\leq C_{\textsf{OPT}}(I_{i})+(h_{1}^{i,1}-o_{1}^{i,1})I_{i}
 \end{equation}
For bounding the second term in the above inequality, we need to introduce the notion of \emph{Residue-Length} as defined below:
 \begin{framed}
 \begin{definition}(Residue-length)
 The $i^{th}$ residue-length $l_i$ is the length of time from the last slot in the previous interval when the \emph{Max-age} user of  the $i^{th}$ interval got served by the \textsf{MA-CSIT} policy, counted up to the beginning of the $i^{th}$ interval.
 \end{definition}
 \end{framed}   
 See Fig.~\ref{up2} for an illustration of the intervals and the residue lengths. 
It is not hard to verify that the difference of the ages of the \emph{Max-age} user under the \textsf{MA-CSIT} policy and \textsf{OPT}  at the beginning of the $i^{th}$ interval can be upper bounded by the residue-length $l_{i}$  \emph{i.e.} $h_{1}^{i,1}-o_{1}^{i,1}\leq l_{i}$. Hence, from Eqn.\ \eqref{int_ineq}, we have the following upper bound: 
\begin{equation}
    C_{\textsf{MA-CSIT}}(I_{i})\leq C_{\textsf{OPT}}(I_{i})+ l_{i}I_{i}.
\end{equation}
Finally, to find an upper bound to the competitive ratio, we need to derive a lower bound of the cost of the \textsf{OPT} policy for each intervals. Note that, after the first time slot of any interval, the \emph{Max-age} user, by definition, encounters consecutive $Bad$ channels. Hence, the cost corresponding to that user under the \textsf{OPT} policy can be lower bounded by $\sum_{k=1}^{I_{i}}k$. 

    During the $i$\textsuperscript{th} interval, the channel corresponding to the user other than the \emph{Max-age} user (\emph{i.e.} user 2) does not become $Good$ after the $(I_{i}-l_{i+1})$\textsuperscript{th} time slot. This fact can be verified from the definition of residue-length. Therefore, the cost for user 2 under \textsf{OPT} for the $i$\textsuperscript{th} interval can be lower bounded as: 
\begin{equation}
    \sum_{k=1}^{I_{i}-l_{i+1}}1+\sum_{k=1}^{l_{i+1}}k=I_{i}-l_{i+1}+\sum_{k=1}^{l_{i+1}}k.
\end{equation}
 Hence, the total AoI cost under the \textsf{OPT} policy (including both users)  for the  $i$\textsuperscript{th} interval can be lower bounded as:
\begin{equation}\label{opt2}
    C_{\textsf{OPT}}(I_{i})\geq \sum_{k=1}^{I_{i}}k + I_{i}-l_{i+1}+\sum_{k=1}^{l_{i+1}}k.
\end{equation}
Summing up the costs over all intervals we have the following bound: 
\begin{equation}\label{bound2}
    \sum_{i}C_{\textsf{MA-CSIT}}(I_{i})\leq \sum_{i}C_{\textsf{OPT}}(I)+\sum_{i}l_{i}I_{i}.
\end{equation}
Substituting the bound from Eqn.\ \eqref{opt2} in the inequality above, we have:
\begin{equation}
\frac{\sum_{i}C_{\textsf{MA-CSIT}}(I_{i})}{\sum_{i}C_{\textsf{OPT}}(I_{i})}\leq 1 +\frac{\sum_{i}l_{i}I_{i}}{\sum_{i}(\sum_{k=1}^{I_{i}}k+I_{i}-l_{i+1}+\sum_{k=1}^{l_{i+1}}k)}.
\end{equation}
Now we use the AM-GM inequality to get $l_{i}I_{i}\leq \frac{l_{i}^{2}}{2}+\frac{I_{i}^{2}}{2}$.\\
 Furthermore, we have
 $\sum_{k=1}^{I_{i}}k+I_{i}-l_{i+1}+\sum_{k=1}^{l_{i+1}}k=\frac{I_{i}(I_{i}+1)}{2}+I_{i}+\frac{(l_{i+1})(l_{i+1}+1)}{2}-l_{i+1}\geq \frac{I_{i}^{2}}{2}+\frac{l_{i+1}^{2}}{2}$. \\
 Hence,
\begin{equation}
    \frac{\sum_{i}C_{\textsf{MA-CSIT}}(I_{i})}{\sum_{i}C_{\textsf{OPT}}(I_{i})}\leq 1+\frac{\sum_{i}\frac{l_{i}^{2}}{2}+\frac{I_{i}^{2}}{2}}{\sum_{i}\frac{I_{i}^{2}}{2}+\frac{l_{i+1}^{2}}{2}}\leq 2,
\end{equation}
where we have used the fact that, by definition $l_1 =0.$
Hence, $\eta^{\textsf{MA-CSIT}}\leq 2$. 
\end{IEEEproof}
The above result should be compared and contrasted with Theorem 3 of \cite{sinha2020optimizing}, which proves an upper limit of $8$ for the competitive ratio of the \emph{Max-Age} policy that operates without CSIT. \\
In the following, we extend the above proof technique for $N=3$ users. The reader will find that although the basic line of analysis remains the same, the details become much more involved in this case.
 \subsection{Competitive ratio of the \textsf{MA-CSIT} policy for $N=3$ users} \label{3user upper bound}
 \begin{framed}
 \begin{prop}
 The competitive ratio of \textsf{MA-CSIT} policy for $N=3$ users is upper bounded as $\eta^{\textsf{MA-CSIT}}\leq \frac{8}{3}.$
 \end{prop}
 \end{framed}
 
\begin{figure}
\centering

\includegraphics[scale=0.35]{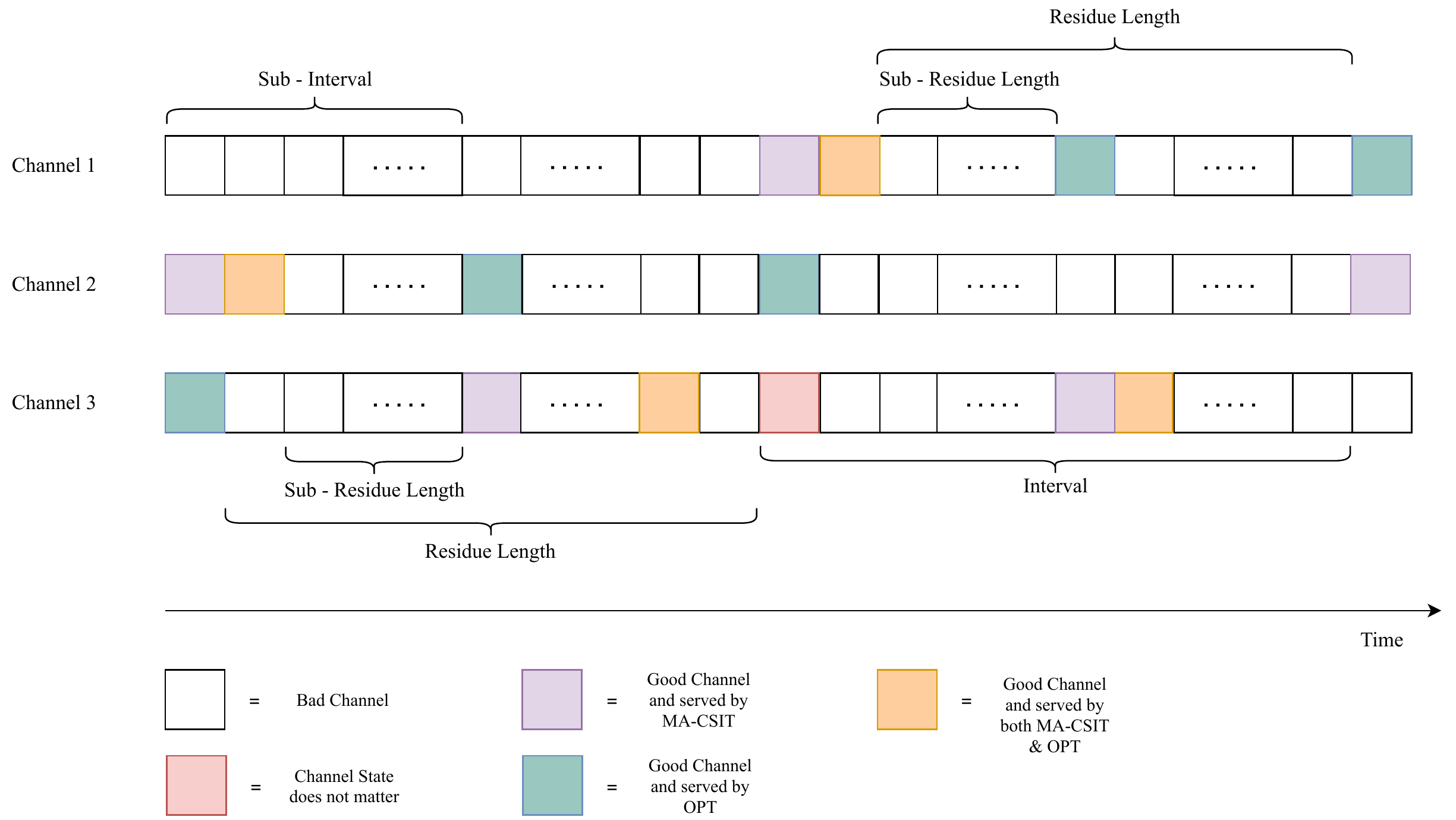}  
\caption{Illustration of residue length and interval construction for 3 users. Channel $i$ refers to the  state of the channel  corresponding to user $i$.}\label{up3}
\end{figure}

\emph{Proof:} We use the same definition of intervals as in our previous proof.
The difference in the AoI costs incurred by the \textsf{MA-CSIT} and the \textsf{OPT} policies can be expressed as follows:
\begin{align}
    C_{\textsf{MA-CSIT}}(I_{i})-C_{\textsf{OPT}}(I_{i}) &\ =\sum_{t}(h_{1}^{i,t}-o_{1}^{i,t})+\sum_{t}(h_{2}^{i,t}-o_{2}^{i,t})+ \nonumber\\
   &\ \sum_{t}(h_{3}^{i,t}-o_{3}^{i,t}),
\end{align}
where the index in the summation ranges over all slots in the $i$\textsuperscript{th} interval.
Without any loss of generality, let us assume that user 1 is the \emph{Max-age} user for the $i^{th}$ interval under the \textsf{MA-CSIT} policy.  Note that \emph{Lemma} 1
holds for any number of users (hence, for $N=3$ also). This is because whenever the  channel corresponding to the \emph{Max-age} user becomes $Good$, the \textsf{MA-CSIT} policy serves that user immediately and a new interval begins. Thus, the difference of ages of the \emph{Max-age} user under the \textsf{MA-CSIT} and \textsf{OPT} policies remains constant throughout any interval (as both increase linearly throughout an interval). Therefore we can write \begin{align}
    C_{\textsf{MA-CSIT}}(I_{i}) &\ =C_{\textsf{OPT}}(I_{i})+(h_{1}^{i,1}-o_{1}^{i,1})I_{i}+\sum_{t}(h_{2}^{i,t}-o_{2}^{i,t})+ \nonumber\\
    &\ \sum_{t}(h_{3}^{i,t}-o_{3}^{i,t}).
\end{align}
Using the same definition of residue-lengths as before, we can express the above difference as:
\begin{equation}
    C_{\textsf{MA-CSIT}}(I_{i})=C_{\textsf{OPT}}(I_{i})+l_{i}I_{i}+\sum_{t}(h_{2}^{i,t}-o_{2}^{i,t})+\sum_{t}(h_{3}^{i,t}-o_{3}^{i,t}).
\end{equation}
We denote the time slot when the Max-age user of the $i$\textsuperscript{th} interval got served by the \textsf{MA-CSIT} policy as $T_{l_{i}}$.\\
 To upper bound the quantity $\sum_{t}(h_{2}^{i,t}-o_{2}^{i,t})+\sum_{t}(h_{3}^{i,t}-o_{3}^{i,t})$ we introduce the notion of \emph{sub-intervals}, which form a partition of the intervals. The formal definition of a sub-interval is given below:
  
 \begin{framed}

 \begin{definition}(Sub-interval)
 Within an interval, a new sub-interval is said to begin  when the \textsf{MA-CSIT} policy serves the $2^{nd}$ Max-age user among the three users.
 \end{definition}
 \end{framed}
Let, $J_{i}$ denotes the number of sub-intervals in the $i$\textsuperscript{th} interval.\\
We define the $j^{th}$ \emph{Sub-Residue length} of the $i^{th}$ interval, $l^{i}_{j}$ as follows:
\begin{framed}
\begin{definition} (Sub-Residue length)
The $j^{th}$ sub-residue length of  the $i^{th}$ interval (denoted by $l^{i}_{j}$) is the time-elapsed since the last time slot when the $2$\textsuperscript{nd} Max-age user of the $j$\textsuperscript{th} sub-interval of the $i$\textsuperscript{th} interval got served by the \textsf{MA-CSIT} policy, up to the beginning of the $(j+1)$\textsuperscript{th} sub-interval.
\end{definition}
\end{framed}
We illustrate the notion of sub-intervals and sub-residue lengths in Fig.~\ref{up3}. Note that, the above definition is analogous to the definition of intervals. 

In every sub-interval, the age of user which has the least age under \textsf{MA-CSIT} policy would be always upper bounded by age of that user under the \textsf{OPT} policy. It directly follows from \emph{Lemma} 2, discussed in the previous section. Hence, we have the following upper bound:
\begin{equation}\label{upper3}
    C_{\textsf{MA-CSIT}}(I_{i})\leq C_{\textsf{OPT}}(I_{i})+l_{i}I_{i}+\sum_{j}l^{i}_{j}I^{i}_{j} 
\end{equation}
where $I^{i}_{j}$ refers to the length of the $j^{th}$ sub-interval of the $i^{th}$ interval, and the index $j$ runs over all sub-intervals of the $i$\textsuperscript{th} interval. 

Next, we proceed to lower bound the cost incurred by the \textsf{OPT} policy during the $i$\textsuperscript{th} interval. 

\paragraph{Lower bounding the cost of the Max-age user} Since, the $i^{th}$ interval continues until the channel corresponding to the \emph{Max-age} user becomes $Good$, the cost incurred by the \emph{Max-age} user (\emph{i.e.} user 1 in this case) 
under the \textsf{OPT} policy is lower bounded by
\begin{equation}\label{opt-max-age}
    \sum_{k=1}^{I_{i}}k\geq \frac{I_{i}^{2}}{2}
\end{equation}
\paragraph{Lower bounding the cost of the $2$\textsuperscript{nd} Max-age user }
The cost incurred by the $2^{nd}$ Max-age user under the \textsf{OPT} policy during the $j^{th}$ sub-interval is $\sum_{k=1}^{I^{i}_{j}}k$. This is true because for the entire duration of the $j$\textsuperscript{th} sub-interval, the channel corresponding to the $2^{nd}$ Max-age user remains $Bad$. 
\paragraph{Lower bounding the cost of the $3$\textsuperscript{rd} \emph{Max-age} user  }
Following the definition of the sub-residue lengths, the quantity $l_{j+1}^{i}$ denotes the last time slot when the \textsf{MA-CSIT} policy serves the $2$\textsuperscript{nd} Max-age user of the $(j+1)$\textsuperscript{th} sub-interval, counted from the beginning of the $(j+1)$\textsuperscript{th} sub-interval. On the $(j+1)$\textsuperscript{th} sub-interval, the $3$\textsuperscript{rd} Max-age user of the $j$\textsuperscript{th} sub-interval becomes the $2$\textsuperscript{nd} Max-age user. Hence, for the last $l_{j+1}^{i}$ time slots of  the $j$\textsuperscript{th} sub-interval, the cost of the $3$\textsuperscript{rd} \emph{Max-age} user of the $j$\textsuperscript{th} sub-interval under the \textsf{OPT} policy is given by $\sum_{k=1}^{l_{j+1}^{i}}k$.\\
Thus, the cost under the \textsf{OPT} policy during the $j$\textsuperscript{th} sub-interval of the $i$\textsuperscript{th} interval, excluding the cost of the \emph{Max-age} user is lower bounded by:
\begin{equation*}
    \sum_{k=1}^{I^{i}_{j}}k+\sum_{k=1}^{I_{j}^{i}-l_{j+1}^{i}}1+\sum_{k=1}^{l_{j+1}^{i}}k \geq \frac{(I_{j}^{i})^{2}}{2}+\frac{(l_{j+1}^{i})^{2}}{2}, \forall 1\leq j \leq J_i-1.
\end{equation*} 
For the last sub-interval of the $i$\textsuperscript{th} interval, the cost incurred by 3\textsuperscript{rd} Max-age user under the \textsf{OPT} policy is lower bounded by $\sum_{k=1}^{I_{J_{i}}^{i}}1$ (since at the $J_{i}$\textsuperscript{th} sub-interval sub-residue length does not exist). Hence the cost incurred by the 2\textsuperscript{nd} Max-age and the 3\textsuperscript{rd} Max-age user under the \textsf{OPT} at the $J_{i}^{th}$ sub-interval is lower bounded by 
\begin{equation}
    \sum_{k=1}^{I_{J_{i}}^{i}}k+\sum_{k=1}^{I_{J_{i}}^{i}}1\geq \frac{(I_{J_{i}}^{i})^{2}}{2}
\end{equation}\\
Finally, summing up the cost over all sub-intervals in the $i$\textsuperscript{th} interval, we get the following lower bound to the cost incurred by the 2\textsuperscript{nd} Max-age and the 3\textsuperscript{rd} Max-age user under the \textsf{OPT} policy:
\begin{equation}\label{subinterval_lower}
 \frac{(I_{J_{i}}^{i})^{2}}{2}+\sum_{j=1}^{J_{i}-1}(\frac{(I_{j}^{i})^{2}}{2}+\frac{(l_{j+1}^{i})^{2}}{2})
\end{equation}
where $l_{J_{i}}^{i}$ is the sub-residue length of the last sub-interval of the $i$\textsuperscript{th} interval and $I_{J_{i}}^{i}$ is the length of the last sub-interval of the $i$\textsuperscript{th} interval.\\
There are three scenarios depending on the values $m^{th}$ residue length $\forall m\in\{2,3,...\}$  \emph{i.e.} $l_{m}$ can take, 
\begin{itemize}
     \item Case 1- $l_{m}\leq I_{m-1}$ ,
     \item  Case 2- $I_{m-2}+I_{m-1}>l_{m}>I_{m-1}$,
    \item Case 3- $l_{m}\geq I_{m-2}+I_{m-1}$.
\end{itemize}
\paragraph*{Case 1} Consider the first scenario where $l_{m}\leq I_{m-1}$ $\forall m\in \{2,3....\}$. Now consider the Max-age user of $(i+1)^{th}$ interval.
The \textsf{MA-CSIT} policy serves the Max-age user  $l_{i+1}$ time slots before the beginning of the $(i+1)^{th}$ interval \emph{i.e} at $T_{l_{i+1}}^{th}$ slot. The \textsf{OPT} policy can serve the Max-age user twice after $T_{l_{i+1}}^{th}$ time slot. Since $l_{i+1}\leq I_{i}$, the \textsf{OPT} policy can serve the Max-age user once at the beginning of $J_{i}^{th}$ sub-interval and next at the beginning of $(i+1)^{th}$ interval. So, the $l_{i+1}$ time slots can be divided into two parts. The first part refers to  the sub-residue length of the last sub-interval \emph{i.e} $l_{J_{i}}^{i}$ and the next one refers to the length of final sub-interval \emph{i.e.} $I_{J_{i}}^{i}$. Hence we have 
\begin{equation}
    I_{J_{i}}^{i}+ l_{J_{i}}^{i}=l_{i+1}
\end{equation}
Let $a_{i}$ denotes the first part of $l_{i}$ and $b_{i}$ refers to the second part. 
So for the above case we have 
\begin{align}\label{ai}
    a_{i+1}=l^{i}_{J_{i}}\\
    b_{i+1}=I^{i}_{J_{i}}\label{bi}
\end{align}
Hence the lower bound of the \textsf{OPT} policy at Eq.~\eqref{subinterval_lower} can be rewritten as 
\begin{align}\label{subinterval_final_lower}
    \frac{b_{i+1}^{2}}{2}+\sum_{j=1}^{J_{i}-2}(\frac{(I_{j}^{i})^{2}}{2}+\frac{(l_{j+1}^{i})^{2}}{2})+\frac{(I_{J_{i}-1}^{i})^{2}}{2}+ \frac{a_{i+1}^{2}}{2}
\end{align}
Since $a_{i+1}=l^{i}_{J_{i}}$ and $b_{i+1}=I^{i}_{J_{i}}$, we can rewrite Eqn.\ \eqref{upper3} as:
\begin{equation}\label{numerator_3user}
    C_{\textsf{MA-CSIT}}(I_{i})\leq C_{\textsf{OPT}}(I_{i})+l_{i}I_{i}+\sum_{j=1}^{J_{i}-1}(l^{i}_{j}I^{i}_{j}) + a_{i+1}b_{i+1}
\end{equation}
Summing the costs over all intervals from Eq.~\eqref{numerator_3user} and using the lower bound of the \textsf{OPT} policy for the Max-age user of Eq.~\eqref{opt-max-age} and the lower bound for the 2nd Max-age and the 3rd Max-age user of Eq.~\eqref{subinterval_final_lower} we get the following bound:
\begin{align}\label{ma-csit3}
   &\ \frac{\sum_{i}C_{\textsf{MA-CSIT}}(I_{i})}{\sum_{i}C_{\textsf{OPT}}(I_{i})}  \leq 1 +\nonumber\\
  &\    \frac{\sum_{i}(l_{i}I_{i})+\sum_{j=1}^{J_{i}-1}(l^{i}_{j}I^{i}_{j})+a_{i+1}b_{i+1})}{\sum_{i}(\frac{I_{i}^{2}}{2}+ \frac{b_{i+1}^{2}}{2}+\sum_{j=1}^{J_{i}-2}(\frac{(I_{j}^{i})^{2}}{2}+\frac{(l_{j+1}^{i})^{2}}{2})+\frac{(I_{J_{i}-1}^{i})^{2}}{2}+ \frac{a_{i+1}^{2}}{2})}
\end{align}
Using the AM-GM inequality, we have   $l_{i}I_{i}\leq \frac{l_{i}^{2}}{2}+\frac{I_{i}^{2}}{2}$, $\sum_{j=1}^{J_{i}-1}(l^{i}_{j}I^{i}_{j})\leq \sum_{j=1}^{J_{i}-1}(\frac{(l_{j}^{i})^{2}}{2}+\frac{(I^{i}_{j})^{2}}{2})$ and $a_{i+1}b_{i+1}\leq \frac{a_{i+1}^{2}}{2}+\frac{b_{i+1}^{2}}{2}$.
Hence, from the above, we get
\begin{align}
   &\ \frac{\sum_{i}C_{\textsf{MA-CSIT}}(I_{i})}{\sum_{i}C_{\textsf{OPT}}(I_{i})}\leq 1 +\nonumber\\
   &\ \frac{\sum_{i}( \frac{l_{i}^{2}}{2} +\frac{I_{i}^{2}}{2}+\sum_{j=1}^{J_{i}-1}(\frac{(l_{j}^{i})^{2}}{2}+\frac{(I^{i}_{j})^{2}}{2})+\frac{a_{i+1}^{2}}{2}+\frac{b_{i+1}^{2}}{2})}{\sum_{i}(\frac{I_{i}^{2}}{2}+ \frac{b_{i+1}^{2}}{2}+\sum_{j=1}^{J_{i}-2}(\frac{(I_{j}^{i})^{2}}{2}+\frac{(l_{j+1}^{i})^{2}}{2})+\frac{(I_{J_{i}-1}^{i})^{2}}{2}+ \frac{a_{i+1}^{2}}{2})}
\end{align}
Combining  above equations we get
\begin{align}
    \frac{\sum_{i}C_{\textsf{MA-CSIT}}(I_{i})}{\sum_{i}C_{\textsf{OPT}}(I_{i})} &\ \leq   2 +\nonumber\\
    &\ \frac{\sum_{i}( l_{i}^{2})}{\sum_{i}({I_{i}}^{2}+2\sum_{j=1}^{J_{i}-1}(\frac{(I^{i}_{j})^{2}}{2}+\frac{(l^{i}_{j+1})^{2}}{2})+b_{i+1}^{2}+a_{i+1}^{2})}
\end{align}
Lower bounding the sub-interval lengths and the sub-residue lengths by zero, from the above,  we have 
\begin{equation}
    \frac{\sum_{i}C_{\textsf{MA-CSIT}}(I_{i})}{\sum_{i}C_{\textsf{OPT}}(I_{i})}\leq 2 +\frac{\sum_{i}( l_{i}^{2} )}{\sum_{i}({I_{i}}^{2} +b_{i+1}^{2}+a_{i+1}^{2})}
\end{equation}

Since, $a_i+b_i=l_i,$ using the Cauchy-Schwartz inequality, we have $a_i^2+b_i^2 \geq l_i^2/2, \forall i.$ Hence, the RHS of the above equation can be further upper bounded as below:
\begin{eqnarray}\label{fin_bd}
\frac{\sum_{i}C_{\textsf{MA-CSIT}}(I_{i})}{\sum_{i}C_{\textsf{OPT}}(I_{i})}\leq 2 + \frac{\sum_{i}( l_{i}^{2} )}{\sum_{i}({I_{i}}^{2} +\frac{l_{i+1}^{2}}{2})}
\end{eqnarray}
We have $l_{i}\leq I_{i-1}$.
Note that the RHS of Eqn.\ \eqref{fin_bd} is monotonically increasing for $l_{i}\geq 0$.
Hence, we can upper bound the RHS of equation \eqref{fin_bd} by substituting $l_{i}=I_{i-1}$. Therefore, we get 
\begin{equation}
  \frac{\sum_{i}C_{\textsf{MA-CSIT}}(I_{i})}{\sum_{i}C_{\textsf{OPT}}(I_{i})}\leq 2+ \frac{\sum_{i}( I_{i-1}^{2} )}{\sum_{i}({I_{i}}^{2} +\frac{I_{i}^{2}}{2})}\leq \frac{8}{3}
\end{equation}
Please see the Appendix  section  for the proof of Case 2 and Case 3.
Hence, for all values of $l_{i}$, we get $ \frac{\sum_{i}C_{\textsf{MA-CSIT}}(I_{i})}{\sum_{i}C_{\textsf{OPT}}(I_{i})}\leq \frac{8}{3}$ which implies $\eta^{MA-CSIT}\leq \frac{8}{3}$.
\section{Simulation Results} \label{sims}
In this section we provide two particular channel configurations for 2 users and 3 users scenario to show the tightness of the bound provided in the \ref{2user-upper bound} and \ref{3user upper bound} sections.
\subsection{$N=2$ users case}
\begin{figure}
\centering
\includegraphics[scale=0.2]{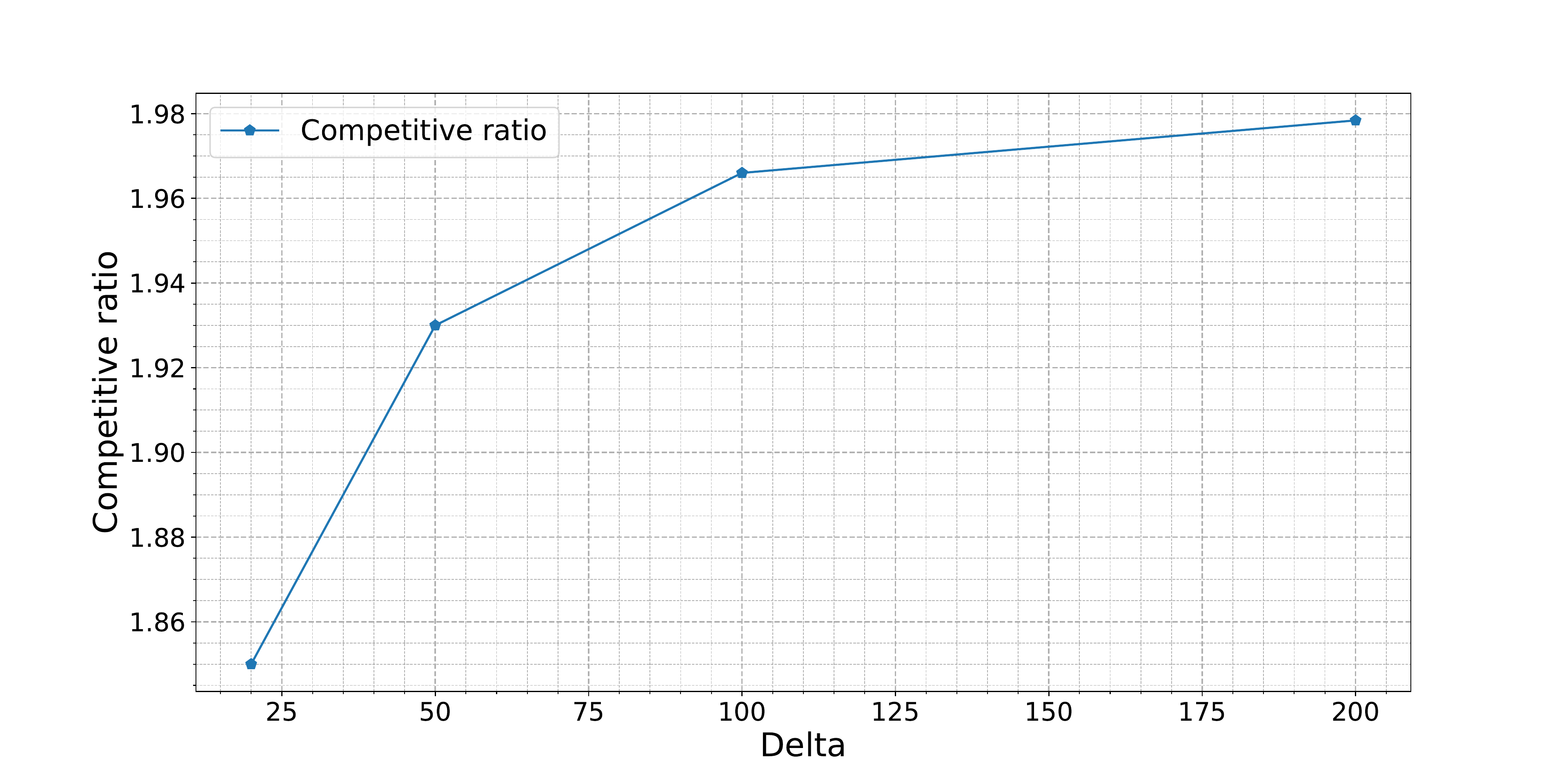}  
\caption{AoI cost comparison between \textsf{MA-CSIT} policy and  \textsf{OPT}  policy for 2 users}\label{comp2}
\end{figure}

Consider the following channel state sequence for 2 users where the whole sequence is divided into intervals of fixed length $\Delta$ where $\Delta$ is even. At the beginning of every interval the channel corresponding to the user 1 is \textsf{Good} and the other channel is \textsf{Bad}. For the next $\frac{\Delta}{2}-2$ slots both channels remain \textsf{Bad}. Next, at the $\frac{\Delta}{2}^{th}$ slot both the channels become \textsf{Good}. After that, at the $(\frac{\Delta}{2}+1)^{th}$ slot, the channel corresponding to user 2 remains \textsf{Good} but other channel becomes \textsf{Bad}. For the next $\frac{\Delta}{2}-2$ slots both channels remain \textsf{Bad} and finally at the $\Delta^{th}$ slot both channels become \textsf{Good}. 
In Fig.~\ref{comp2} the AoI cost ratio between the \textsf{MA-CSIT} and the \textsf{OPT} policy for this particular channel state configuration has been plotted. It can be seen as interval length grows the cost ratio approaches 2, while in section \ref{2user-upper bound} we showed that for 2 user case the competitive ratio for the \textsf{MA-CSIT} policy is upper bounded by 2. 

\subsection{$N=3$ users case}
\begin{figure}
\centering

\includegraphics[scale=0.2]{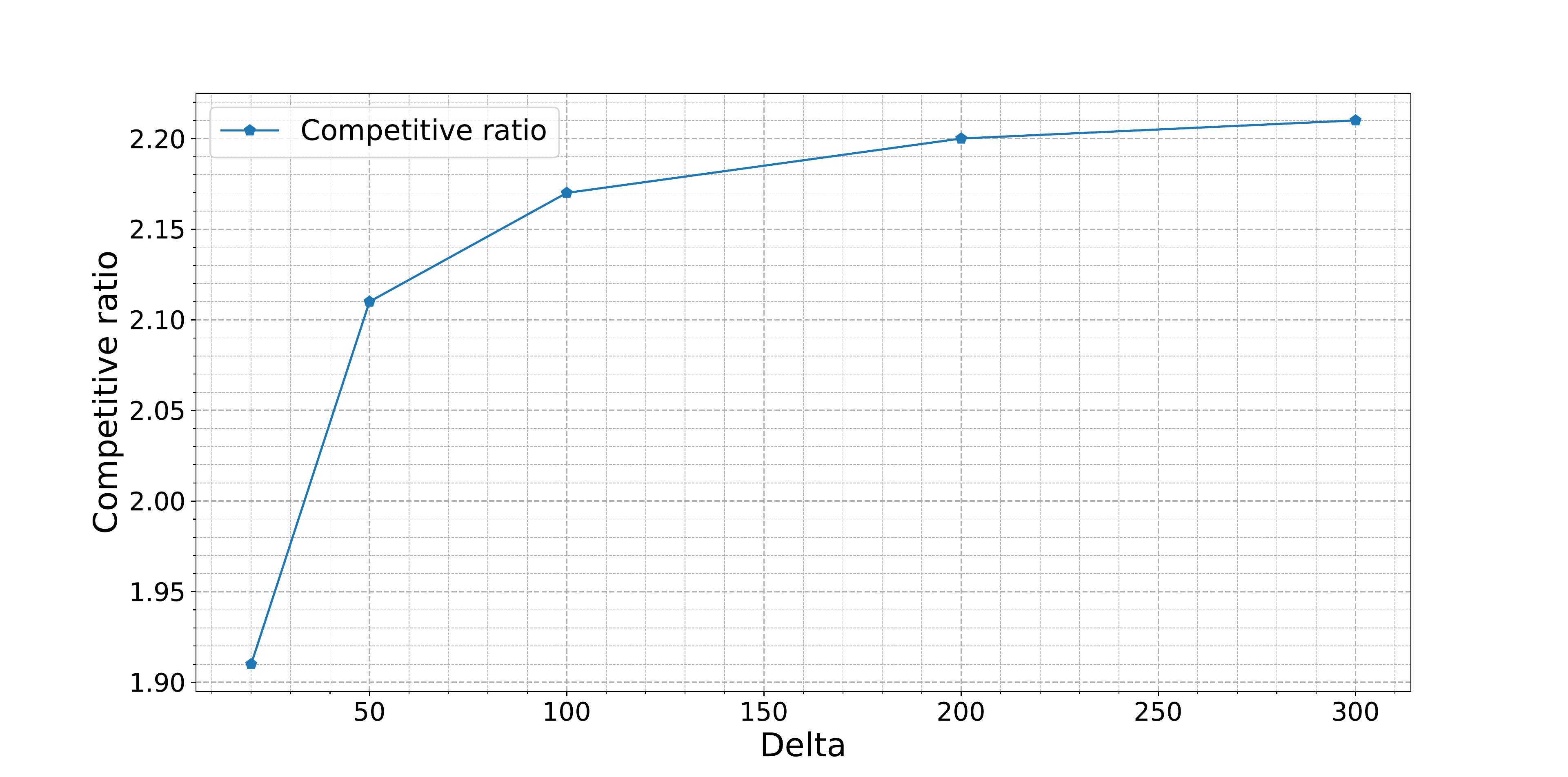}  
\caption{AoI cost comparison between \textsf{MA-CSIT} policy and  \textsf{OPT}  policy for 3 users}\label{comp3}
\end{figure}

In this case, we consider the interval length $\Delta$ to be multiple of 6. Here we mention at which time slot  the channels corresponding to the users become \textsf{Good}.   At the first time slot of the interval the channels corresponding to user 1 and 2 are only \textsf{Good}. For $2^{nd}$ and $3^{rd}$ time slots the channels corresponding user 1 and user 3 remain \textsf{Good} respectively. At the $(\frac{\Delta}{6})^{th}$ time slot, the channels corresponding to user 1 and user 3 become \textsf{Good} and at the next time slot, the channel corresponding to the user 1 only remains \textsf{Good}. After that, at the $(\frac{\Delta}{3}+1)^{th}$ slot, the channels corresponding to user 2 and user 3 become \textsf{Good}.
At next two time slots \emph{i.e.} $(\frac{\Delta}{3}+2)^{th}$ and $(\frac{\Delta}{3}+3)^{th}$  slots the channels corresponding to user 2 and user 1 remain \textsf{Good} respectively. Next at the $\frac{\Delta}{2}^{th}$ time slot  the channels corresponding to user 1 and user 2 become \textsf{Good} and at the next time slot, the channel corresponding to user 2 only remains \textsf{Good}. After that at $(\frac{2\Delta}{3}+1)^{th}$ slot, the channels corresponding to user 1 and 3 become \textsf{Good}.
For the next two time slots  \emph{i.e.} $(\frac{2\Delta}{3}+2)^{th}$ and $(\frac{2\Delta}{3}+3)^{th}$  slots  the channels corresponding to user 3 and user 2 remain \textsf{Good} respectively.  Next at $\frac{5\Delta}{6}^{th}$ time slot,  the channels corresponding to user 2 and user 3 become \textsf{Good} and at the next time slot, the channel corresponding to user 3 only remains \textsf{Good}. In all other time slots the users which are not mentioned, the channels corresponding to those users remain \textsf{Bad}. 
 For this particular scenario the AoI cost ratio between the \textsf{MA-CSIT} and the \textsf{OPT} policy has been plotted in Fig.~\ref{comp3}. As $\Delta$ grows, the cost ratio approaches 2.25, while in section \ref{3user upper bound} we showed that for 3 user case the competitive ratio for the \textsf{MA-CSIT} policy is upper bounded by 2.67. 


\section{Comparison between the \textsf{MA-CSIT} and the Max Age policy}

\begin{figure}
\centering

\includegraphics[scale=0.2]{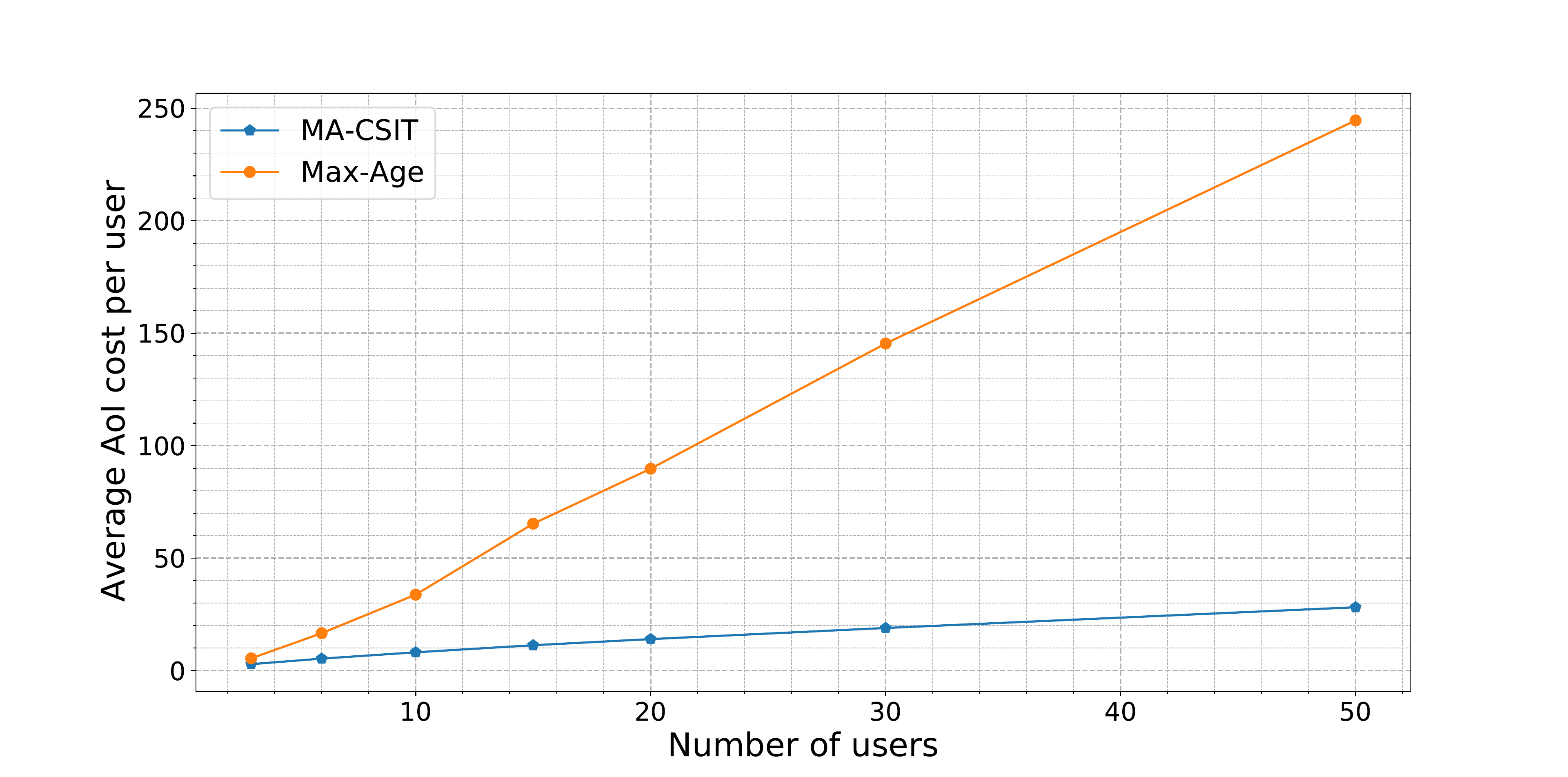}  
\caption{AoI cost comparison between \textsf{MA-CSIT} policy and  Max-Age policy}\label{ma-csit2}
\end{figure}

\begin{figure}
\centering

\includegraphics[scale=0.2]{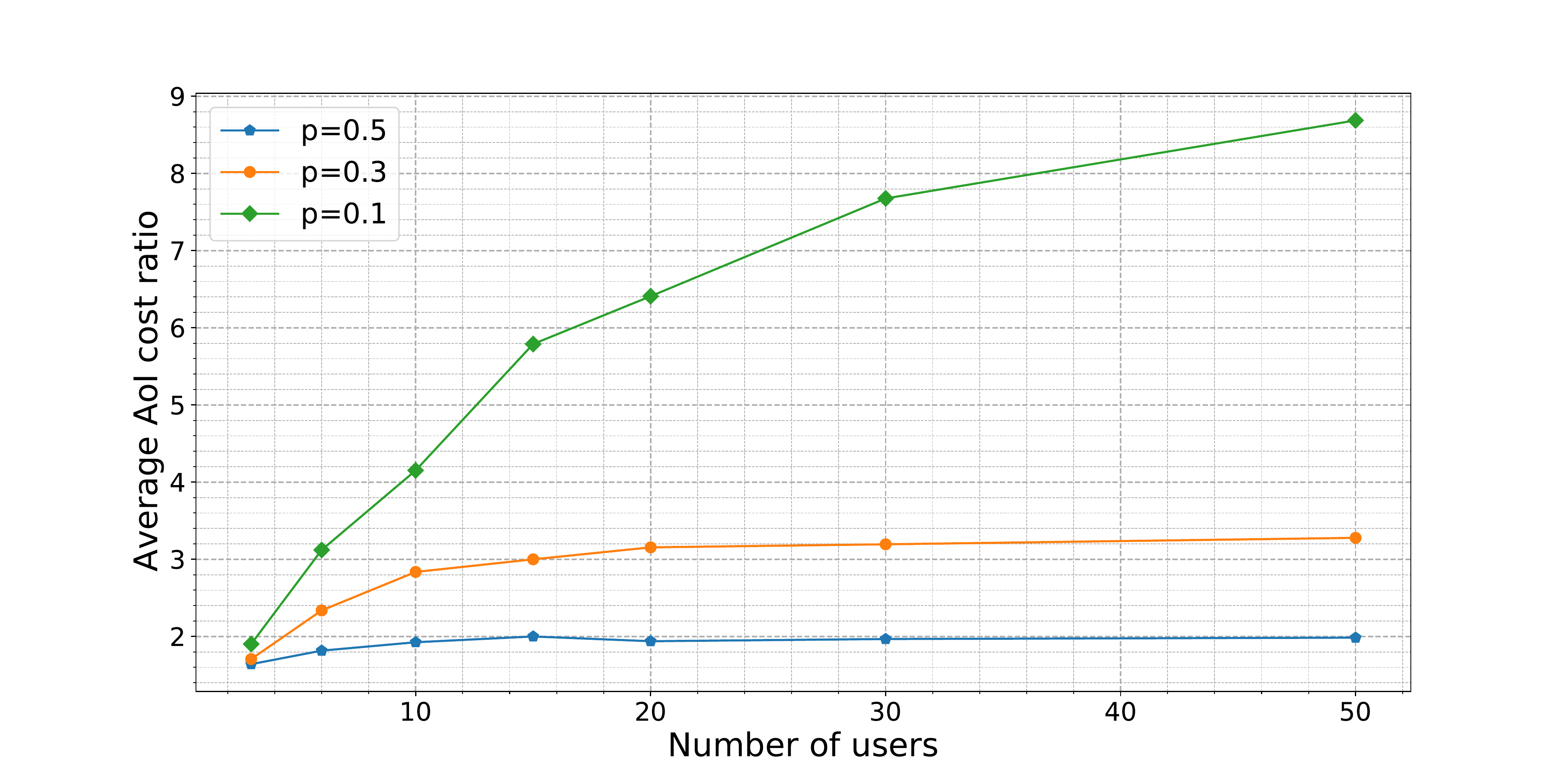}  
\caption{Ratio between AoI cost of \textsf{MA-CSIT} policy and that Max-Age policy}\label{ratio}
\end{figure}

In this section we provide numerical results to show the advantage of having CSIT. 
Through simulations we compare the performance of \textsf{MA-CSIT} policy and the Max-Age policy \cite{ISIT2020Sinha,sinha2020optimizing} which does not have CSIT. In this case we consider the channel states corresponding to each user to be independent and identically distributed.
\begin{figure}
\centering

\includegraphics[scale=0.2]{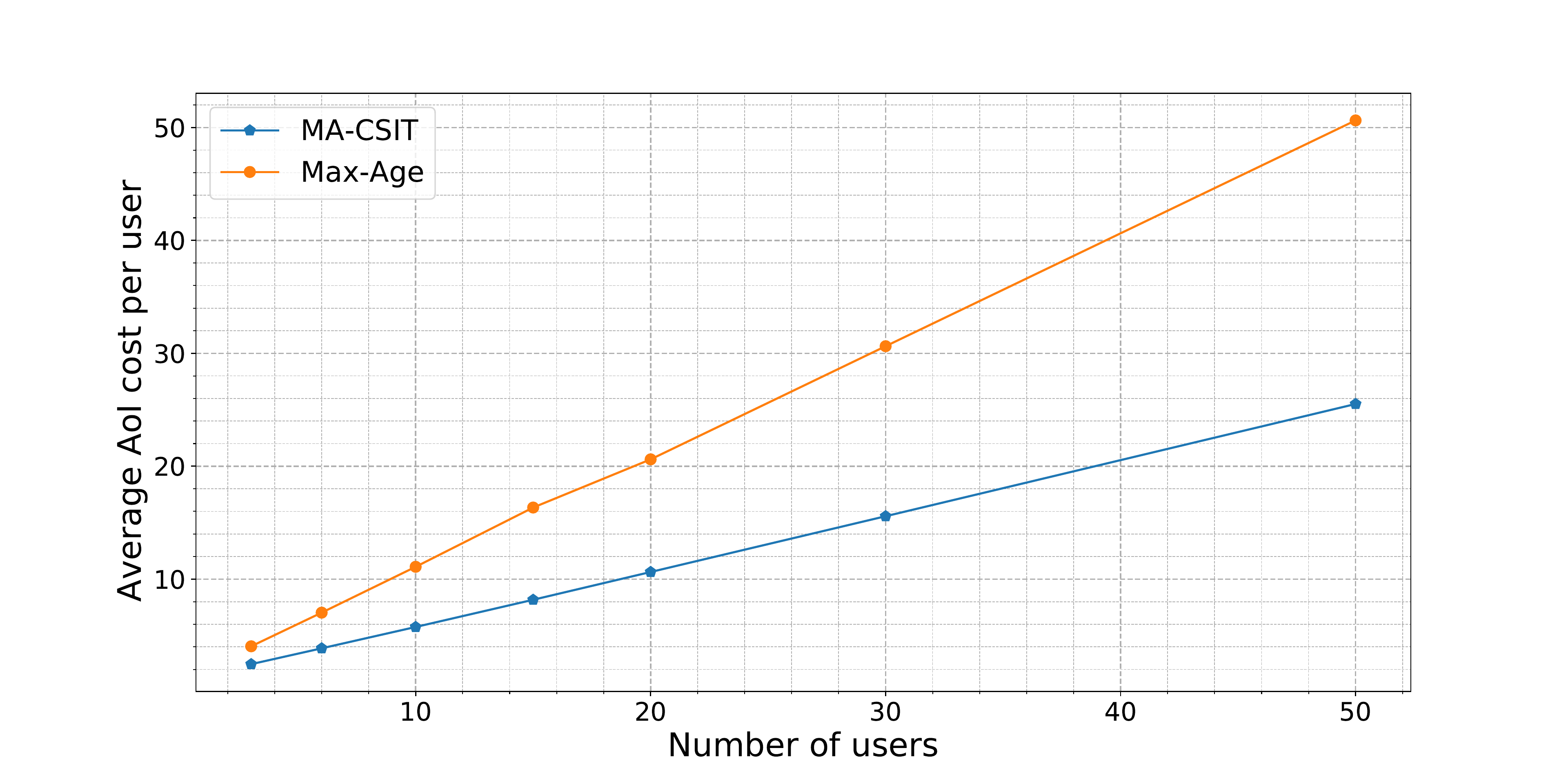}  
\caption{AoI cost comparison between \textsf{MA-CSIT} policy and  Max-Age policy}\label{ma-csit}
\end{figure}
Consider the channel corresponding to each user can be \textsf{Good} with a probability $p$.
\begin{figure}
\centering

\includegraphics[scale=0.2]{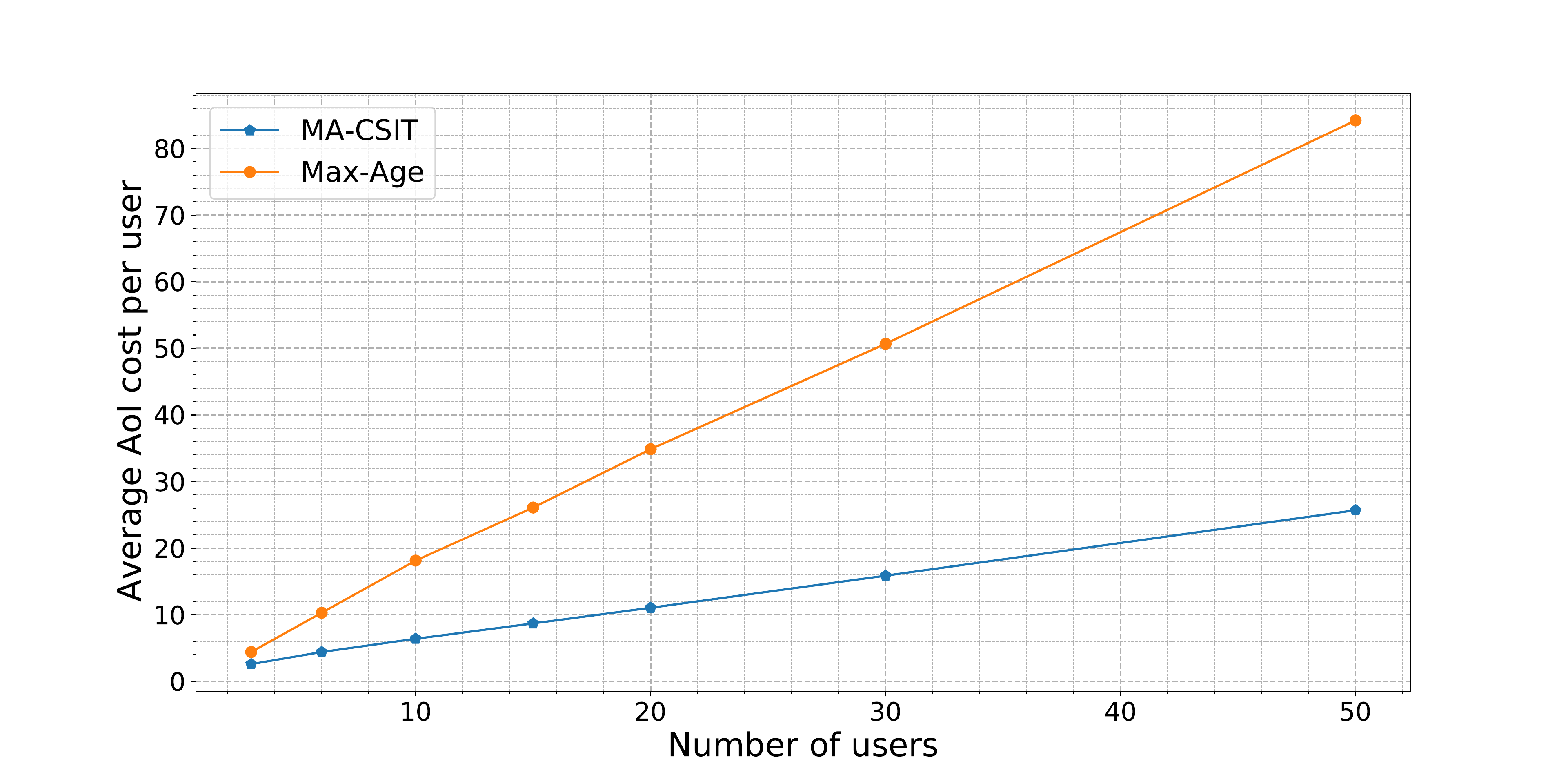}  
\caption{AoI cost comparison between \textsf{MA-CSIT} policy and  Max-Age policy}\label{ma-csit1}
\end{figure}
In Fig.~\ref{ma-csit}, Fig.~\ref{ma-csit1} and Fig.~\ref{ma-csit2}, the time averaged AoI cost ($AoI_{avg}(T)$) for \textsf{MA-CSIT} policy and Max Age policy when $p=0.5$, $p=0.3$ and $p=0.1$ have been plotted respectively. 
In Fig.~\ref{ratio},  the ratio between the average AoI cost of Max Age policy and that of \textsf{MA-CSIT} policy for these three cases has been shown. From the plots we can conclude that when the policy is equipped with CSIT the performance improves drastically.

\section{Conclusion} \label{conclusion}
The paper investigates the fundamental limits of Age-of-Information for static users over adversarial environments when the scheduling policy is assumed to know  the CSIT at the current slot. Theoretically we provide upper bound on the competitive ratio when the number of users is either 2 or 3. Through simulations, we showed that the greedy scheduling policy performs substantially  better over adversarial setting when the policy is equipped with the channel state information at the current slot. Finding an upper bound on the competitive ratio for arbitrary number of users is an interesting open problem.
\section{Acknowledgement} \label{ack}
This work is partially supported by the grant IND-417880 from
Qualcomm, USA and a research grant from the Govt. of India for the potential Center-of-Excellence \emph{Intelligent Networks} under the IoE initiative. The computational results
reported in this work were performed on the AQUA Cluster at
the High Performance Computing Environment of IIT Madras.

\clearpage

\bibliographystyle{IEEEtran}

\bibliography{comsnets.bib}
\clearpage
\section{Appendix}\label{appen}
\paragraph*{Case 2: $I_{m-2}+ I_{m-1}>l_{m}>I_{m-1}$ }
In this case, before the $i^{th}$ interval the last time slot at which the \textsf{MA-CSIT} policy can serve the Max-age user of $i^{th}$ interval would lie somewhere at the $(i-2)^{th}$ interval. At that time slot that user has the least age under the \textsf{MA-CSIT} policy. We denote that time slot as $T_{l_{i}}^{i}$ and the Max-age user of $i^{th}$ interval as $u_{max}^{i}$. 
Next we need to determine the time slots where the \textsf{OPT} serves $u_{max}^{i}$ but the \textsf{MA-CSIT} does not.
\begin{itemize}
    \item After $T_{l_{i}}^{i}$ time slot suppose, the \textsf{OPT} serves $u_{max}^{i}$ at some time slot at $(i-2)^{th}$ interval but the \textsf{MA-CSIT} does not. This is only possible when the 2nd Max-age user and $u_{max}^{i}$ get \textsf{Good} channels.
Since at the $(i-2)^{th}$ interval, $u^{i}_{max}$ has the least age under the \textsf{MA-CSIT} policy, the \textsf{MA-CSIT} serves the 2nd Max-age user. Hence at that time slot $u_{max}^{i}$ becomes the $2^{nd}$ Max-age user and at the $(i-1)^{th}$ interval it will become the Max-age user. But it is not possible, as $u_{max}^{i}$ is the Max-age user of $i^{th}$ interval and same user can not become Max-age user at two consecutive intervals.
\item Another possible scenario is after $T_{l_{i}}^{i}$ the $2^{nd}$ Max-age user at the $(i-2)^{th}$ interval  gets \textsf{Bad} channels constantly and at the beginning of $(i-1)^{th}$ interval the channels  corresponding to both $u_{max}^{i}$ and the 2nd Max-age user  become \textsf{Good} and the \textsf{OPT} serves $u_{max}^{i}$ instead of serving the 2nd Max-age user. In this scenario the \textsf{MA-CSIT} serves the Max-age user and the $2^{nd}$ Max-age user becomes the Max-age user of the $(i-1)^{th}$ interval \emph{i.e.} $u^{i-1}_{max}$. But in this case the \textsf{OPT} policy can serve $u_{max}^{i-1}$ at max once after the user gets served by the \textsf{MA-CSIT} policy which implies
\begin{align}\label{}
    a_{i-1}=l_{J_{i-1}}^{i-1}=l_{i-1}
\end{align}

At $T_{l_{i}}^{i}$ time slot the \textsf{OPT} policy can serve either 2nd Max-age user or the $u^{i}_{max}$. But if the \textsf{OPT} policy serves the $u^{i}_{max}$, then for the rest of the time slots of $(i-2)^{th}$ interval and entire $(i-1)^{th}$ interval the cost difference for that user under the \textsf{OPT} and the \textsf{MA-CSIT} policy remains zero. \\
Suppose the \textsf{OPT} policy serves the 2nd Max-age user. Since after $T_{l_{i}}^{i}$ time slot, both $u^{i}_{max}$ and 2nd Max-age user get \textsf{Bad} channels constantly and $l^{i-1}_{J_{i-1}}\leq I^{i-1}_{J_{i-1}-1}$ , the cost difference between the \textsf{MA-CSIT} and the \textsf{OPT} policy for the users other than the Max-age user for rest of the $(i-2)^{th}$ interval 
\begin{equation}
    l^{i-1}_{J_{i-1}}I^{i-1}_{J_{i-1}}- I^{i-1}_{J_{i-1}-1}I^{i-1}_{J_{i-1}}\leq 0
\end{equation}
 Hence the cost difference between the \textsf{MA-CSIT} policy and the \textsf{OPT} policy  at $(i-2)^{th}$ interval is
 \begin{align}
     C_{\textsf{MA-CSIT}}(I_{i-2})-&\ C_{\textsf{OPT}}(I_{i-2})\leq  \frac{l_{i-2}^{2}}{2} + \nonumber\\
     &\ \frac{I_{i-2}^{2}}{2}+\sum_{j=1}^{J_{i-2}-1}(\frac{(l_{j}^{i-2})^{2}}{2}+\frac{(I^{i-2}_{j})^{2}}{2}
 \end{align}
 For this particular case at the $(i-1)^{th}$ interval there will not be any sub-interval because $u^{i}_{max}$ can not get \textsf{Good} channel at the $(i-1)^{th}$ interval, otherwise the \textsf{MA-CSIT} policy will serve $u^{i}_{max}$ immediately and this will contradict the assumption $l_{i}>I_{i-1}$.
Thus, for $(i-1)^{th}$ interval, we have 
\begin{equation}
    \sum_{j=1}^{J_{i-1}-1}(\frac{(l_{j}^{i-1})^{2}}{2}+\frac{(I^{i-1}_{j})^{2}}{2})=0
\end{equation}
Since the \textsf{OPT} policy serves the $u^{i}_{max}$ at the beginning of $(i-1)^{th}$ interval we have 
\begin{align}
    b_{i}=I_{i-1}
\end{align}
Hence at $(i-1)^{th}$ interval the cost difference between  the \textsf{MA-CSIT} policy and the \textsf{OPT} policy is 
\begin{align}
    C_{\textsf{MA-CSIT}}(I_{i-1})-C_{\textsf{OPT}}\leq l_{i-1}I_{i-1}+a_{i}b_{i}
\end{align}
Since the OPT policy did not serve the 2nd Max-age user at the beginning of $(i-1)^{th}$ interval the cost of $u^{i-1}_{max}$ under OPT policy is lower bounded by $\sum_{k=1}^{l_{i}+a_{i-1}}k\geq \frac{(l_{i}+a_{i-1})^{2}}{2}$.\\
Now consider, $N_{1}=\sum_{k}I_{\{k\neq i-2,i-1\}}(\frac{l_{k}^{2}}{2} +\frac{I_{k}^{2}}{2}+\sum_{j=1}^{J_{k}-1}(\frac{(l_{j}^{k})^{2}}{2}+\frac{(I^{k}_{j})^{2}}{2})+\frac{a_{k+1}^{2}}{2}+
    \frac{b_{k+1}^{2}}{2})$, $
    N_{2} = \frac{l_{i-2}^{2}}{2} +\frac{I_{i-2}^{2}}{2}+\sum_{j=1}^{J_{i-2}-1}(\frac{(l_{j}^{i-2})^{2}}{2}+\frac{(I^{i-2}_{j})^{2}}{2}$
and     $N_{3}=l_{i-1}I_{i-1}+a_{i}b_{i}$.
Also let $D_{1}=\sum_{k}I_{\{k\neq i-2, i-1\}}(\frac{I_{k}^{2}}{2}+\sum_{j=1}^{J_{k}-1}(\frac{(I^{k}_{j})^{2}}{2}+\frac{(l^{k}_{j+1})^{2}}{2})+\frac{a_{k+1}^{2}}{2}+\frac{b_{k+1}^{2}}{2})$, $D_{2}=\frac{I_{i-2}^{2}}{2}+\sum_{j=1}^{J_{i-2}-1}(\frac{(I^{i-2}_{j})^{2}}{2}+\frac{(l^{i-2}_{j+1})^{2}}{2})+\frac{a_{i-1}^{2}}{2}=\frac{I_{i-2}^{2}}{2}+\sum_{j=1}^{J_{i-2}-1}(\frac{(I^{i-2}_{j})^{2}}{2}+\frac{(l^{i-2}_{j+1})^{2}}{2})+\frac{l_{i-1}^{2}}{2}$ (since $a_{i-1}=l_{i-1}$) and $D_{3}=\sum_{m=1}^{l_{i}}m+\sum_{m=1}^{a_{i}}m+\sum_{m=1}^{b_{i}}m$.\\
In the expression of $D_{2}$ the first summation indicates the lower bound on the cost incurred by $u^{i-1}_{max}$ under the \textsf{OPT} policy since it got served by the \textsf{OPT} policy for the last time before $(i-1)^{th}$ interval. The rest two summations refer to the lower bound of  the cost incurred by $u^{i}_{max}$ under the \textsf{OPT} policy since it got served by the \textsf{MA-CSIT} policy before $i^{th}$ interval.
\begin{align}
   \frac{\sum_{i}C_{\textsf{MA-CSIT}}(I_{i})}{\sum_{i}C_{\textsf{OPT}}(I_{i})} &\ \leq  + \frac{N_{1}+N_{2}+N_{3}}{D_{1}+D_{2}+D_{3}} \nonumber\\
   &\ \leq 1+ \nonumber\\
   &\ \frac{N_{1}+ N_{2}+\frac{l_{i-1}^{2}}{2}+\frac{I_{i-1}^{2}}{2}+\frac{a_{i}^{2}}{2}+\frac{b_{i}^{2}}{2}}{D_{1}+ \frac{I_{i-2}^{2}}{2}+\frac{l_{i-1}^{2}}{2}+\frac{l_{i}^{2}}{2}+\frac{a_{i}^{2}}{2}+\frac{b_{i}^{2}}{2}}
   \end{align}
Since $b_{i}=I_{i-1}$ and $l_{i}\geq I_{i-1}$, simplifying above equations we get,
\begin{align}\label{macsit5}
  &\  \frac{\sum_{i}C_{\textsf{MA-CSIT}}(I_{i})}{\sum_{i}C_{\textsf{OPT}}(I_{i})} \leq 2 + \nonumber\\
 &\    \frac{\sum_{k}I_{\{k\neq i-1,i\}}( \frac{l_{k}^{2}}{2})+ \frac{I_{i-1}^{2}}{2} }{\sum_{k}I_{k\neq i-1,i}(\frac{I_{k}^{2}}{2}+\frac{a_{k+1}^{2}}{2}+\frac{b_{k+1}^{2}}{2})+\frac{l_{i}^{2}}{2}+\frac{I_{i-1}^{2}}{2}+\frac{l_{i-1}^{2}}{2}+\frac{a_{i}^{2}}{2}} 
\end{align}
Hence we have 
\begin{align*}
    \frac{\sum_{i}C_{\textsf{MA-CSIT}}(I_{i})}{\sum_{i}C_{\textsf{OPT}}(I_{i})} &\ \leq 2 +\\
    &\ \frac{\sum_{k}I_{\{k\neq i\}}I_{k}^{2}}{\sum_{k}I_{\{k\neq i,i-1\}}(I_{k}^{2}+a_{k+1}^{2}+b_{k+1}^{2})+2I_{i-1}^{2}}\\
    &\ \leq \frac{8}{3}
\end{align*}

\end{itemize}
 
\paragraph*{Case 3} When $l_{i}\geq I_{i-1}+I_{i-2}$, it is easy to check that for $u_{max}^{i}$ cost under \textsf{OPT} will be always greater than the cost under \textsf{MA-CSIT} policy. Hence $ \frac{\sum_{i}C_{\textsf{MA-CSIT}}(I_{i})}{\sum_{i}C_{\textsf{OPT}}(I_{i})}$ is upper bounded by \nicefrac{8}{3}.
\end{document}